\documentclass[aps, twocolumn,floats, showpacs]{revtex4}
\usepackage{amsmath,amssymb,graphicx}
\usepackage{epsfig}
\usepackage{graphicx}
\usepackage{enumerate}

\newcommand{\beq}{\begin{equation}}
\newcommand{\eeq}{\end{equation}}

\newcommand{\bea}{\begin{eqnarray}}
\newcommand{\eea}{\end{eqnarray}}

\begin{document}
\title{Thermal conductance of the Fermi-Pasta-Ulam chains:
Atomic to mesoscopic transition}
\author{Yelena Nicolin and Dvira Segal}
\affiliation{Chemical Physics Theory Group, Department of Chemistry, University of Toronto,
80 Saint George St. Toronto, Ontario, Canada M5S 3H6}

\begin{abstract}
We demonstrate that in the atomic-scale limit
the thermal conductance  $\mathcal K$ of the FPU model and its variants strongly
deviates from the  mesoscopic behavior due to the relevance of contact resistance. As a result,
atomic chains follow $\log \mathcal K = \nu \log T$, where the power law coefficient $\nu$ is
exactly two times larger than the mesoscopic value.
We smoothly interconnect the atomic and mesoscopic limits, and demonstrate that this turnover behavior
takes place in other nonlinear FPU-like  models. %xxx other potentials.
%We also present intriguing results for the thermal conductance of atomic chains very-weakly coupled to the contacts.
Our results are significant for nanoscale applications, manifesting an atomic thermal
conductance with temperature scaling superior to the mesoscopic limit.
\end{abstract}

 % 63.22.+m Phonons or vibrational states in low-dimensional structures and nanoscale materials
 %44.10.+i Heat conduction (see also 66.60.+a and 66.70.+f in transport properties of condensed matter)
 %05.60.-k Transport processes
% 05.45.-a  Nonlinear dynamics and chaos
% %66.70.+f Nonelectronic thermal conduction and heat-pulse propagation in solids; thermal waves (for thermal conduction in metals and alloys, see 72.15.Cz and 72.15.Eb63.20.Ry Anharmonic lattice modes)
%{\today}

\maketitle

%-------------------

% FPU
The Fermi-Pasta Ulam  (FPU) model \cite{FPU} and its variants
provide an ideal test-bed for addressing fundamental issues in
statistical mechanics such as equipartition of energy,  the onset of
chaos in nonlinear dynamical systems, and the validity of
macroscopic laws in low dimensional systems \cite{Chaosrev}. In
particular, the thermal conductivity of the FPU model has been
extensively investigated, demonstrating a breakdown of the
normal-diffusional Fourier's law dynamics \cite{Lepri}. The
underlying question addressed in these %analytical and numerical
studies has typically been whether a one dimensional (1D) chain of
oscillators with a specific force-field can demonstrate equilibrium
and dynamical properties characterizing macroscopic objects. Thus,
simulations were mostly carried out using long chains of $10^3-10^4$
beads, targeting the mesoscopic-macroscopic limit.

Nonetheless, in recent years the thermal conduction properties of {\it nanoscale
junctions} has been of fundamental and practical interest \cite{Majumdar}.
Here, the physical setup includes a nanoscale object, e.g., a nanotube \cite{CNT} or
an alkane molecule \cite{Dlott}, coupled to two (or more) thermal contacts.
As  devices size shrinks, an enormously important question is the applicability of the macroscopic
dynamical laws at the nanoscale. Even in the framework of classical mechanics, contact effects and the constriction geometry
may imply on the onset of new scaling rules.
The interest in nanoscale thermal conduction is driven by various challenges.
One of the major problems in molecular electronics
is junction heating,  %caused by the electron current exciting vibrational modes,
limiting the device stability. % and reproducibility.
Understating heat transport in molecular
systems thus attract considerable attention both experimentally \cite{Majumdar,Dlott,Selzer,DNA}
and theoretically \cite{Kirczenow,SegalH,Diventra,Dicarlo,Galperin}.
Managing thermal transport across interfaces \cite{Keblinski} %e.g. a solid-liquid  or a solid-solid interface
is also important in microelectronic devices \cite{Majumdar},
 biophysical applications \cite{Bio}, and in thermoelectric energy conversion devices \cite{thermoE}.

Motivated  by these challenges we focus here on the {\it
atomic-scale} steady-nonequilibrium FPU model and its variants. As
the effect of the contact (interface) and internal nonlinear
interactions cannot be trivially
separated, a complex-new dynamics is revealed at different
temperature domains for both weak and strong system-bath interaction
strengths.
%
%alpha FPU other models%%.
%We simulate heat transfer in representative anharmonic chains,
%the FPU-$\beta$ and the quartic model (or the infinite-temperature FPU),
%and obtain new results for the thermal conductance in the atomic limit.
%
Specifically, using numerical simulations we demonstrate that when
the contact resistance controls the dynamics, e.g., for extremely
short chains or at relatively low temperatures (yet above the
harmonic-anharmonic transition), the conductance of FPU-type systems
follows a $\mathcal K\propto T^{\nu}$ power law,
 where $\nu=1/2$ for the $\beta$-FPU model.
%This dynamics is also obtained for intermediate chains at low enough temperatures,
This behavior stands in a sharp contrast to the
mesoscopic $\beta$-FPU limit where the value $\nu=1/4$ is obtained,
%the scaling $\mathcal K \propto T^{1/4}$ is obtained,
%i.e., the power is exactly two times smalles,
as expected from the phenomenological Debye theory.
We justify our results within the effective phonon theory \cite{Albasio,Cai},
%valid even for strong nonlinear systems \cite{Albasio,Cai},
and further interconnect the atomic and mesoscopic limits, demonstrating a
smooth turnover of the dynamics with increasing chain size and
temperature.
Our simulations also pinpoint on the critical temperature where contact resistance is secondary to the
bulk thermal resistance.

% discuss the turnover temperture for harmonic-anhamonic transition and for contact-bulk

%-------------------------------------------
% MODEL
We consider a 1D lattice of $N$ atoms whose Hamiltonian reads
\bea
&&H=\sum_{i=1}^{N} \frac{p_i^2}{2} +  g_2 \Omega^2 V_2  +g_4\beta V_4, \\
&&V_2= \frac{1}{2}\sum_{i=0}^{N}(x_{i+1}-x_{i})^2 ; \,\,
V_4= \frac{1}{4}\sum_{i=0}^{N} (x_{i+1}-x_{i})^4.
\nonumber
\label{eq:H}
\eea
Here $x_i$ is the displacement from the equilibrium position of the
$i-th$ particle, $x_0$ and $x_{N+1}$ are the fixed boundaries,
$\beta$ measures the strength of nonlinear interactions, and $g_s$
($s=2,4)$ are Boolean variables taking the values of 0 or 1. In what
follows we will consider three models: (i) A harmonic model ($H_2$)
with $g_2=1$, $g_4=0$, (ii) a quartic model ($H_4$) with $g_2=0$,
$g_4=1$, and (iii) the $\beta$-FPU model ($H_F$) with $g_2=1$,
$g_4$=1. The quartic model is introduced here to facilitate
identifying the dynamics of the FPU model at high temperatures.

The direct way to determine the thermal conductance $\mathcal{K}$ of a 1D chain
is to couple the left and right ends of the system with two thermal baths
at temperatures $T_L$  and $T_R$, respectively. In our simulations
we use Langevin thermostats, with the motivation to simulate
real experiments on the nanoscale, where a small molecule connects to macroscopic solids
\cite{Dlott,DNA}. In such systems the contact resistance is an unavoidable issue.
We note that in mesoscopic scale simulations Nose-Hoover thermostats are typically
used \cite{Li1, Li2, Aoki}. However, these thermostats do not correctly reproduce
 the weak-coupling regime \cite{Lepri}, which is of interest here as well.
The chain's particles obey the following equations of motion,
\bea
%\ddot x_1&=&-\frac{\partial H}{\partial x_i}-\gamma_L \dot x_1 +F_L
%\nonumber\\
\ddot x_i&=&-\frac{\partial H}{\partial x_i}-\left( \gamma_R \dot x_i -F_R\right) \delta_{i,N}
-\left( \gamma_L \dot x_i -F_L \right) \delta_{i,1}.
\nonumber\\
%\ddot x_{i}&=& -\frac{\partial H}{\partial x_i}, \,\,\,\,\,\,\, i\neq 1,N
\label{eq:EOM}
\eea
where $\gamma_{L,R}$ is the coupling strength between the system and the $L,R$ bath, and
the Gaussian thermal white noises obey the fluctuation-dissipation relation ($k_B=1$),
$\langle F_{L,R}(t)\rangle=0$,
$\langle F_{L,R}(t) F_{L,R}(0)\rangle=2\gamma_{L,R}T_{L,R}\delta(t)$.
The long-time heat flux can be calculated between every two sites as
the force exerted by the $i-th$ particle on the $i+1$ oscillator,
%(check sign),
$J_i=-\langle \dot x_i[g_2\Omega^2(x_{i+1}-x_i)+ g_4 \beta(x_{i+1}-x_i)^3 ] \rangle$
%
%\bea
%J_i=-\langle \dot x_i\partial V(x_{i+1}-x_i)/ \partial x_i \rangle,
%\eea
%
where the average reflects the time and ensemble average, performed after the system has reached steady-state.
The thermal conductance is defined as,
%
%\bea
$\mathcal{K}= \frac{1}{\Delta T }\sum_{i=1}^{N-1} J_i/(N-1)$,
%\eea
%
where $\Delta T=T_L-T_R$.
%
%-----------------------------------------

\begin{figure}[htbp]
%\hspace{2mm}
{\hbox{\epsfxsize=70mm \epsffile{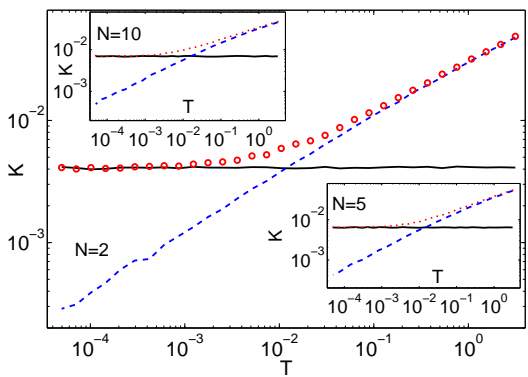}}}
\caption{
Thermal conductance of atomic FPU chains (circles or dots), a quartic chain (dashed line)
and an harmonic chain (full line) as a function of the average temperature $T=(T_L+T_R)/2$.
$\gamma_L=\gamma_R=0.6$, $\Omega^2=0.01$, $\beta=0.01/4$.
The main plot displays data for $N$=2. The lower (upper) panel displays data for $N=5$ ($N=10$).
}
\label{FigN2}
\end{figure}

\begin{figure}[htbp]
%\hspace{2mm}
%{\hbox{\epsfxsize=80mm \epsffile{figN20.ps}}}
{\hbox{\epsfxsize=70mm \epsffile{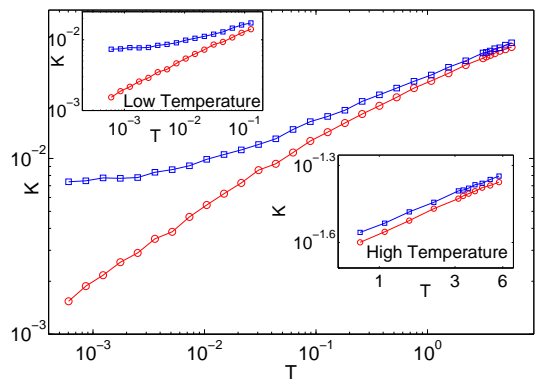}}} \caption{Power law
behavior at low and high temperatures.
%Thermal conductance of atomic FPU chains ($\square$) and a quartic chain (o).
$\gamma_L=\gamma_R=0.6$, $\Omega^2=0.01$, $\beta=0.01/4$, $N=20$,
an FPU chain ($\square$), a quartic chain (o).
The  upper and lower panels zoom on data in the low and high temperature limit respectively.
}
\label{FigN20}
\end{figure}

%------------------------------------------
In Fig. \ref{FigN2} we present the thermal conductance of relatively
short chains with $N=2,5,10$, for the harmonic, quartic, and the FPU
models in the strong friction limit using $\Omega=0.1$,
$\gamma=\gamma_{L,R}=0.6$, $\beta=0.0025$, and  $T_L=0.8T_R$. The
following observations can be made: (i) Disregarding anharmonic
interactions, the conductance does not depend on temperature, (ii)
the quartic model shows a power-law dependence,  $\mathcal{K}\sim
T^{1/2}$,  (iii) the FPU model smoothly interconnects the harmonic
and quartic results, with a crossover temperature around $10^{-2}$,
and (iv) for long chains the power law coefficients at low and high
temperatures are distinct. The upper inset ($N=10$) demonstrates
this effect, to be discussed in more details in Fig. \ref{FigN20}.
%
% Harmonic part.
We can explain these results as follow. For a 1D {\it harmonic}  chain
it can be analytically  shown  \cite{Lebowitz} that the average stationary thermal flux is
given by
\bea
J=\frac{\Omega^2  }{2\gamma}\phi(1) \Delta T,
\label{eq:Jh}
\eea
where $\phi(j)=\frac{\sinh(N-j)\alpha}{\sinh N\alpha}$ and
$e^{-\alpha}=1+z/2 - \sqrt{z+z^2/4}$, $z=\Omega^2/\gamma^2$. In the
strong coupling limit, $z\rightarrow 0$, and we obtain $\alpha\sim
\sqrt z$, resulting in $\phi(1)\sim 1/2$ for $N=2$ and $\phi(1)\sim
1$ for $N\gg 1$. Thus, the strong coupling harmonic conductance  reduces to
\bea
\mathcal{K}_{N=2}= \frac{\Omega^2}{4\gamma}; \,\,\,\,\,\,\,\ \mathcal{K}_{N\rightarrow \infty}= \frac{\Omega^2}{2\gamma}.
\label{eq:Kc}
\eea
In the opposite weak-coupling limit ($z\rightarrow \infty$) we get  $\alpha \sim \ln z$, $\phi(1) \sim 1/z$,
and $\mathcal{K}=\gamma/2$. irrespective of chain size.
In the present case, Eq. (\ref{eq:Kc}) predicts the values 0.0041, 0.0064, 0.0070
for the $N=2,5,10$ chains (respectively). Our simulation data perfectly agree with these numbers.
%in a  perfect agreement with the simulation data of 0.0041, 0.0064 and 0.0069.
At low temperatures, since the dominant contribution to the transport
comes from the quadratic potential,
the FPU model obeys (\ref{eq:Kc}), alike a pure harmonic system.

% High temp limit
We turn next to the low temperature quartic model.
Here a power-law dynamics is observed with $\mathcal K\sim T^{0.5}$
($N$=2). In the high temperature regime ($T>10^{-2}$)
both FPU and quartic models yield $\mathcal K\sim T^{0.4}$.
These observations stand in a sharp contrast to other results observing a $T^{1/4}$
dependency \cite{Li1,Li2,Aoki}. Note however that these works were
concerned with  the thermal conductivity of periodic or long
anharmonic chains, where contact effects were negligible or
non-existing. In contrast, in short atomic systems contact
resistance plays a crucial role \cite{DNA,Lepri,Keblinski}. We
consider its influence next.

Assuming that the total thermal resistance of an atomic-scale chain is given by the sum $R_T=R_M +R_C$,
where $R_M$ is the bulk molecular resistance,
and $R_C$ is the contact resistance, one finds that the total conductance
satisfies $\mathcal K_T =\mathcal K_M \mathcal K_C/(\mathcal K_M + \mathcal K_C)$,
where $\mathcal K_n=1/R_n$; $n=T,M,C$.
Since the molecular resistance increases with size
(not necessarily linearly), we can generally assume that
for short chains $K_T\sim \mathcal K_C$, while for long systems
$\mathcal K_T\sim \mathcal K_M$.
Next we estimate separately the contact conductance $\mathcal K_C$ and the
molecular contribution $\mathcal K_M$.

%{\it Contact thermal conductance.}
It is well justified that a sort of a Virial theorem holds for nonlinear
potentials, resulting in normal mode spectra which are
pseudo-harmonic for both the quartic potential and the FPU model
\cite{Albasio}, with the dispersion relation
\bea
\tilde \omega_k ^2 = \alpha \Omega^2 \omega_k^2.
\label{eq:scale}
\eea
Here the integer $k$ counts the normal modes,
$\omega_k=2\sin(k\pi/N)$, and the renormalization factor satisfies
\bea
\alpha= g_2+ g_4 \frac{2\beta}{\Omega^2} \frac{\langle V_4\rangle_H } {\langle V_2\rangle_H}.
\label{eq:alphaA}
\eea
The average over the potential energy is taken with respect to either the $H_2$,
$H_4$ or the $H_F$ Hamiltonian.
It is remarkable that the factor $\alpha$ does not depend on the wavenumber $k$.
The above relation is valid for systems at equilibrium. In our simulations
 only a small temperature bias is applied, thus
we pose the ansatz that (\ref{eq:scale}) holds in the nonequilibrium regime as well, i.e.,
the equipartition relation, $T=\alpha \Omega^2 \omega_k^2 \langle q_k^2 \rangle_{T}$, is valid
close to equilibrium,  where averages are taken with respect to the mean temperature $T=(T_L+T_R)/2$.
Here $q_k$'s are the normal modes of a 1D harmonic chain.
Our numerical results, validating (\ref{eq:s2}) below, justify this conjecture \cite{comment2}.
Calculating the thermal averages in (\ref{eq:alphaA}),
in the high temperature regime the dynamics of the FPU model
follows the behavior of the
quartic model, while at low $T$ it coincides with the harmonic dynamics,
\bea
\alpha \sim
\begin{cases}
 \frac{\beta}{\Omega^2} T^{\frac{1}{2}}  & {\rm High \,} T\\
 1 & {\rm Low \,} T.
\end{cases}
\label{eq:alpha}
\eea
Since for short chains the boundary resistance dominates heat
transfer, the conductance is assumed to follow Eq. (\ref{eq:Kc}),
%
%\bea
$\mathcal K_C \propto \alpha \Omega^2/\gamma$,
%\eea
%
corrected by the renormalization factor (\ref{eq:alpha}). This leads
to the following scaling law
\bea
&& \mathcal K_C   \propto \alpha \sim
\begin{cases}
T^{\frac{1}{2}} & {\rm High \,} T \\
T^0  &{\rm Low\,} T.
\end{cases}
\label{eq:s2}
\eea
This behavior is significant since a a macroscopic-phenomenological
theory provides a very different scaling law for the FPU model and
its variants: Within the Debye formula the {\it heat conductivity}
can be written as $\kappa=\sum_k c_k v_k^2 \tau_k$, where $c_k$,
$v_k$ and $\tau_k$ are the specific heat, phonon velocity, and the
phonon relaxation time  respectively.
%The $k$ index counts the phononic modes.
Based on this expression, the following relation can be derived
\cite{Li1, Li2}, $\kappa \propto \sqrt{\alpha}/\epsilon$, where
$\alpha$ is the same factor as in Eq. (\ref{eq:alpha}), and
$\epsilon=\langle V_4\rangle_H/  \langle g_4V_4+g_2V_2\rangle_H$. In
the high temperature limit this results in the following proportion,
valid for both the quartic and the FPU models,
%\bea
$\kappa\propto \sqrt{\alpha} \propto T^{\frac{1}{4}}$.
%\eea
%
Assuming that the temperature gradient is linear in our system (besides the contact drop \cite{Aoki}),
see \cite{comment}, we retrieve the bulk conductance
\bea
\mathcal K_M \propto \sqrt{\alpha} \sim  T^{\frac{1}{4}}.
\label{eq:s4}
\eea
We thus conclude that $\mathcal K_C\propto T^{\nu}$ whereas
$\mathcal K_M\propto T^{\nu/2}$. While in the low temperature limit
and for short chains we expect $\mathcal K_C$ to dominate the
overall conductance, in the opposite limit the bulk resistance
dominates, and eventually the scaling $\mathcal K\sim T^{1/4}$
emerges.
%Calculating $\alpha$ for the lower half ($T_a<0.007$) yeilds $\alpha=0.49$, while for the upper part $T_a>0.02$ $\alpha=0.43$.

\begin{figure}[htbp]
%\hspace{2mm}
{\hbox{\epsfxsize=75mm \epsffile{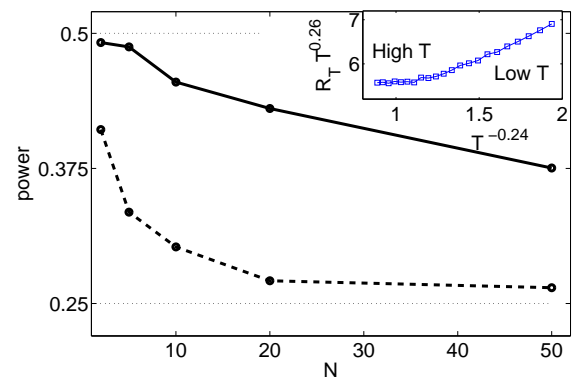}}} \caption{Power low behavior
of the quartic potential as a function of chain size in the high temperature regime, $T>1$ (full line), low
temperature regime $T<0.01$ (dashed).
Data was obtained by studying the slope of
log-log plots of conductance vs. temperature, with
parameters as in Figs. \ref{FigN2} and \ref{FigN20}.
%In the high $T$ regime a clear
%transition from  $\nu=0.5$ to $\nu=0.25$ is observed.
%manifesting the reduced importance of contact effects for long chains.
Inset: Thermal resistance times $T^{0.26}$ for $N$=50,
marking the dominance of bulk conductance for $T>0.7$.}
\label{FigS}
\end{figure}

We present next data for a longer chain of $N$=20, see Fig.
\ref{FigN20}. Focusing on the quartic model (circles), it is evident
that the low-$T$  and the high-$T$ power-law coefficients
deviate. Specifically, at low $T$ the quartic model follows
$\mathcal K \sim T^{0.41}$ (upper panel) while at high $T$, $\mathcal
K \sim T^{0.26}$ (lower panel). The high $T$ values are
characteristic for the FPU model as well ($\square$). Furthermore,
we can systematically extract the power-law coefficients as a
function of chain size at different temperature ranges, see Fig.
\ref{FigS}. While at low $T$ the power-law behavior is approximately
fixed, $\mathcal K \sim T^{\nu}$ with  $\nu \sim 0.4-0.5$, at high
$T$ a clear transition to the mesoscopic result is obtained with a
power $\nu \sim 0.25$ for $N>20$ at $T>1$.
For convenience the slopes were estimated using the
quartic-potential data.
%but the FPU model reduces to this behavior as well.

We can further identify the critical temperature $T_c$ where bulk conductance dominates contact effects.
We conjecture that $R_T\sim a T^{-1/2}+ b T^{-1/4}$, where the first (second) term reflects
the contact (bulk) resistance, and $a$ and $b$ are ($N$-dependent) constants.
A plot of $R_T T^{1/4}$ vs. $T^{-1/4}$ should thus become flat above a critical temperature.
For $N=50$, the inset of Fig. \ref{FigS} demonstrates this turnover at the value 1.1, which translates into
 $T_c\sim 0.7$.
 We estimate that at this value the phononic mean free path is significantly
shorter than the molecular length. The coefficients $a$ and $b$ are roughly the slope of the linear line and the
asymptotic (constant) value, respectively.
% mean free path.

%-----------------------

%--------------------

% other models - justify with the virial theorem
The divergence of the bulk conductance from the contact conductance,
given by the scaling laws (\ref{eq:s2}) and (\ref{eq:s4}) for the
$\beta$-FPU models, should take place in other models of confining
potentials. In particular, we study the thermal conductance of the 2-6 potential, $H=H_T +  g_2
\Omega^2 V_2 +g_4\beta V_6$, with $V_6= \frac{1}{6}\sum_{i=0}^{N}
(x_{i+1}-x_{i})^6$. Here $H_T$ is the kinetic energy and $V_2$
is the harmonic term, same as in (\ref{eq:H}).
In the high temperature limit we can estimate $\alpha$ by \cite{Albasio}  % defined for the $\beta$-FPU model in (\ref{eq:alpha}), by
%
%\bea
$\alpha\sim \frac{\langle V_6\rangle_H}{\langle V_2 \rangle_H }= \frac{\sqrt{\pi}  }{3 \Gamma(5/6)}  T^{2/3}$,
%\eea
%
where $\Gamma$ is the Gamma function. Short chains, controlled by the contact resistance,
are expected to follow $\mathcal K \propto \alpha \sim T^{2/3}$, beyond the harmonic-anharmonic transition.
%in the high temperature regime.
%The  pure $V_6$ model, discarding the harmonic part will show this behavior at low $T$ as well.
In contrast, long chains should obey $\mathcal K \propto
\sqrt{\alpha}\sim T^{1/3}$. Our numerical simulations have verified
this behavior (not shown).
Similarly, studying the $\alpha$-FPU
model, we obtained a power law behavior with $\nu=1/3$
%$\mathcal K \propto T^{1/3}$
for short chains, contrasting the mesoscopic $\nu=1/6$ value \cite{Li2}.
%$\mathcal K\propto T^{1/6}$ \cite{Li2}.

%---------------------
\begin{figure}[htbp]
%\hspace{2mm}
{\hbox{\epsfxsize=70mm \epsffile{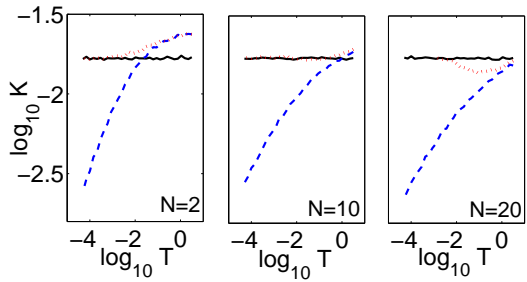}}}
\caption{Conductance of the harmonic chain (full), quartic chain (dashed)
and the $\beta$-FPU chain (dotted) for various chains in the weak coupling limit
$\gamma<\Omega$,
 $\Omega^2=0.01$, $\gamma_L=\gamma_R=0.05$, $\beta=0.01/4$. }
 \label{Figga}
 \end{figure}
%----------------

% small gamma
As our scaling argument relays on the validity of Eq. (\ref{eq:Jh})
for nonlinear systems, renormalizing the phonon spectrum, we further
test the applicability of this relation in the weak coupling limit,
$\gamma \ll \Omega$. In this parameter range this expression predicts
that the conductance of both harmonic and FPU-like systems should be
constant, independent of temperature and size. % $\mathcal K \sim \gamma/2$.
%since this expression does not depend on the phonon frequency.
Our numerical simulations, Fig. \ref{Figga}, indeed confirm this expectation:
At low temperatures both the harmonic potential and the
$\beta$-FPU model yield $\mathcal K\sim \gamma/2 =0.0025$. In the
high temperature limit small deviations from this value disclose
contributions to energy transfer beyond the effective-phonon theory
\cite{Aoki}.
It is also worth noting the intricate dynamics for
$N$=20. As expected,  with increasing temperature  the FPU
model nicely interpolates the harmonic to the quartic limit.
However, quite interestingly, since the conductance of the
anharmonic chain lies {\it below} the harmonic value, the FPU model
demonstrates a decrease of $\mathcal K$ with increasing $T$ around
the harmonic-anharmonic transition, see the rightmost panel in Fig.
(\ref{Figga}).

%------------------------------------------
In summary, we presented here new results for the temperature
scaling of the conductance of anharmonic chains with confining
potentials. In the atomistic limit we found that the contact thermal
resistance controls the junction conductance, leading to a dynamics
significantly distinct from the mesoscopic limit where  bulk
conductance dominates.
These results are significant fundamentally, demonstrating a
new atomic limit with enhanced thermal conductance for the  celebrated FPU model, driven by interface effects.
From the practical point of view our results deliver an encouraging message for nanoscale applications, manifesting
that the combination of contact effects and nonlinearity could be  beneficial for thermal transport, leading to an enhanced conductivity
in comparison to the macroscopic limit.

\noindent \emph{Acknowledgments.} This research has been supported
by NSERC.
%--------------------------------------------------------------

%-----------------------------


\begin{thebibliography}{32}
\bibitem{comment}
Note the distinction between the thermal conductance $\mathcal K=J/\Delta T$
and the thermal conductivity $\kappa=-J/\nabla T$.
We have numerically verified  that at high enough $T$, the temperature gradient
is constant along both FPU and quartic junctions,  $\nabla T \sim \Delta T/N$.
besides the finite jumps at the boundaries.
Thus, the temperature scaling of $\kappa$ and $\mathcal K$ should be the same.
This is also justified using the  profile
$T(x)=T_L\left\{   1- \left[  1-\left(\frac{T_R}{T_L}\right)^{1-\gamma}\right]\frac{x}{N}  \right\}^{1/(1-\gamma)}$
valid for systems manifesting the power law conductivity $\kappa\sim T^{-\gamma}$ \cite{Aoki}.
For  $\Delta T\ll T_L,T_R$ one can show that
$T(x)\sim T_L(1+\Delta T/T_L \frac{x}{N})$, thus the conductance and conductivity scale similarly with
temperature.
\bibitem{comment2}
The simulations were consistent for $T_L=\eta T_R$, for various $\eta$ close to 1.
\end{thebibliography}
\end{document}